\documentclass[aps,prb,twocolumn,showpacs,amsmath,amssymb,superscriptaddress]{revtex4-1}

\usepackage{color}
\usepackage{appendix}

\usepackage{graphicx}
\usepackage{dcolumn}
\usepackage{amsmath} 
\usepackage{amssymb}

\usepackage{xcolor}
\usepackage{epstopdf}

\begin{document}
\title {Breathing mode of a quantum droplet in a quasi-one-dimensional dipolar Bose gas}

\author{E. Orignac}
\affiliation{Univ Lyon, Ens de Lyon, CNRS, Laboratoire de Physique, F-69342 Lyon, France}
\author{S. De Palo}
\affiliation{CNR-IOM-Democritos National 
UDS Via Bonomea 265, I-34136, Trieste, Italy}
\affiliation{Dipartimento di Fisica Teorica, Universit\`a Trieste, Strada Costiera 11, I-34014 Trieste, Italy}
\author{L. Salasnich}
\affiliation{Dipartimento di Fisica e Astronomia ``Galileo Galilei'' and QTech,
Universit\`a di Padova, via Marzolo 8, I-35131 Padova, Italy}
\affiliation{Istituto Nazionale di Fisica Nucleare, Sezione di Padova, via Marzolo 8, I-35131 Padova, Italy}
\affiliation{Istituto Nazionale di Ottica del Consiglio Nazionale delle Ricerche, Unit\`a di Sesto Fiorentino, 
via Carrara 1, I-50019 Sesto Fiorentino (Firenze), Italy}
\author{R. Citro}
\affiliation{Dipartimento di Fisica ``E. R. Caianiello'', Universit\`a
  degli Studi di Salerno and CNR-SPIN, Via Giovanni Paolo II, I-84084 Fisciano (Sa), Italy}
\affiliation{Istituto Nazionale di Fisica Nucleare, Gruppo Collegato di Salerno, Via Giovanni Paolo II, I-84084, Fisciano (Sa), Italy}
  
\begin{abstract}
 We investigate the breathing mode and the stability of a quantum droplet in a tightly trapped one-dimensional dipolar gas of bosonic atoms. When the droplet with a flat-top density profile is formed, the breathing mode frequency scales as the inverse of the number of atoms in the cloud. This is  straightforwardly derived within a phenomenological hydrodynamical approach and confirmed  
using both a variational method based on a generalized Gross-Pitaevskii action functional and the sum-rule approach. 
We extend our analysis also to the presence of axial confinement showing the effect of the trap on the density profile and therefore on the breathing mode frequency scaling.
Our analysis confirms the stability of the quantum droplet against the particle emission when the flat-top density profile is observed. Our results can be used as a guide to the experimental investigations
of collective modes to detect the formation of quantum droplets in quasi-one-dimensional dipolar gases.
\end{abstract}

\date{\today}
\maketitle
\section{Introduction}
\label{sec:intro}

In the last years the research activity on quantum droplets in atomic gases has become very 
intense and fruitful. The achievement of Bose-Einstein condensation of lanthanide atoms with
large magnetic moments (like $^{164} Dy$ or $^{166} Er$) \cite{Chomaz2023}, where the
competition of short and long-range interactions results in a variety of phases, allowed the
observation of self-bound droplets, leading to the subsequent realization of dipolar 
supersolids in cigar-shaped potentials\cite{pfau2016,pfau_2016,ferlaino_2016, boettcher2019}.
For this geometry it has been shown, both theoretically and experimentally, that increasing the
harmonic confinement along the direction of the dipoles, the ground state of the system passes
from a single-droplet to multiple-droplet phase. For a strictly one-dimensional dipolar bosonic gas
predictions for the formation of self-bound droplet are available \cite{oldziejewski_strongly_2019,depalo2021,depalo2022} 
however they haven't been observed experimentally yet. Beyond dipolar gases, droplets have also been observed in attractive bosonic mixtures \cite{tarruell2018a,Semeghini2018,derrico2019}. 

Among the collective excitations, the breathing mode frequency has been demonstrated to be a basic and reliable tool for investigating the nature of the
state in the atomic clouds. The crossover from a non-interacting gas described by 
the one-dimensional mean-field Thomas-Fermi regime to a Tonks-Girardeau gas and then to a super-Tonks-Girardeau gas, has been experimentally and theoretically studied for bosonic cesium atoms\cite{Haller_science_2009} characterized by contact type of interactions. 
Large number of theoretical investigations have then been devoted to the study of the spectrum of excitations in dipolar gases, as for the three-dimensional case where droplets are formed\cite{Santos_2016} or in the cigar-like configuration \cite{Baille2017, Baille2018}.

In this paper we investigate the response of a quantum droplet to a longitudinal compression in a quasi-one-dimensional dipolar Bose gas in a regime describing the one-dimensional tubes experimentally achievable as in Ref.~\onlinecite{Lev_therm_2018,Lev2020}.  
This configuration differs from the case of cigar-shaped traps\cite{derrico2019} where a transverse component in the atomic cloud is still relevant. Thanks to a phenomenological approach that relies on a specific density profile of a droplet, the {\it flat-top}, 
we find that the breathing mode scales as $N^{-1}$, with $N$ the number of the particle in the cloud. We then estimate the breathing mode for a
{\it real} droplet density profile using a generalized Gross-Pitaevskii (GGPE) approach that allows to take  explicitly into account the contributions of density gradients. We then test the findings for the breathing mode estimates using the sum-rule approach, in the limit
of zero trapping confinement. We find that when there is a self-bound droplet with a flat-top density profile the $N^{-1}$ scaling is still fulfilled.
We rely on the equation of state obtained using our previous Variational Bethe-Ansatz that encodes the contribution of both contact and 
dipolar interactions on equal footing\cite{depalo2020,depalo2021,depalo2022,depalo2023} for the quasi-one dimensional dipolar gas for a
range of density and scattering lengths from experimental configuration\cite{Lev_therm_2018}. Results for the stability of the droplet against the particle emission when the trapping potential is released are also discussed. Moreover we find a hierarchy between the breathing mode frequency estimated via the GGPE and the one using the sum-rule as a result of the Holder inequality. 
Finally we extend the previous approaches to the presence of the harmonic oscillator confinement and discuss the changes in the density profile of the droplet and the corresponding breathing frequency. 

The paper is organized as follows:  In Sec.\ref{sec:breathing_mode}
the different approaches for estimating the breathing mode are presented. Together with a brief description of the sum-rule approach, the phenomenological approach (in Sec. \ref{sec:phenomenological}) and generalized Gross-Pitaevskii (in Sec. \ref{sec:ggpe}) approach are discussed.

This Section also contains the extension to the  case of a finite trapping potential.
In Sec.~\ref{sec:results} we test the predictions obtained in Sec.\ref{sec:phenomenological} and Sec. \ref{sec:ggpe} against the sum-rule estimates of the breathing mode frequency.
Finally in Sec.\ref{sec:conclusions} we give conclusions and an outlook. 
{In the Appendices we give technical details for 
a quantum treatment using the Dirac-Frenkel variational
principle (App.~\ref{app:tdva}), for sum rules and H\"{o}lder
inequality (App.~\Ref{app:holder}), for the deformation of the
droplet due to harmonic trapping (App.~\ref{app:deformation}) and 
finally for a discussion of the properties of the ground state
wavefunction of the Generalized Gross-Pitaevskii equation (App.~\Ref{app:ggpe-gs}).}

\section{Breathing mode for a droplet}
\label{sec:breathing_mode}
We can evaluate the frequency of the lowest longitudinal compressional oscillation of the cloud in a longitudinal harmonic trap  
\begin{equation}
H_{\text{long}} = \frac 1 2 m \omega_{ho}^2 \sum_j x_j^2, 
\label{eq:ham_long}
\end{equation}
with {$x_j$ the longitudinal coordinates of the atoms and} 
$\omega_{ho}$ the longitudinal trapping frequency, using a sum-rule approach~\cite{menotti_stringari}. It allows to compute the breathing mode frequency
from ground-state density profiles alone, since 
the breathing mode $\Omega_b$ is obtained as the response of the droplet to a change of the trap frequency $\omega_{ho}$,   
\begin{equation}
\Omega^2_b= -2 \langle \sum^N_{i=1} x^2_i\rangle \left[\frac{\partial \langle \sum^N_{i=1} x^2_i\rangle}{\partial \omega^2_{ho} }\right]^{-1}. 
\label{eq:sum_rule}
\end{equation}

The breathing mode can also be expressed in terms of $m_1/m_{-1}$, where $m_{1,-1}$ are the moments of order plus or minus one of the square radius operator (see App.~\ref{app:holder} for definitions). 
For a 1D Bose gas with repulsive contact interaction, the squared ratio of breathing mode to trapping frequency starts from $4$ in the ideal gas limit\cite{hu_2014,choi_2015,chen_2015,gudyma2015}, smoothly decreases to $3$ in the mean-field Gross-Pitaevskii regime and then increases back to 4 in the Tonks-Girardeau limit\cite{test}, when the system can be seen as impenetrable bosons\cite{girardeau_bosons1d}. We can observe the same evolution in the quasi-1D dipolar gas with repulsive contact interaction\cite{depalo2021}. 
 When the effective interaction is attractive and the system evolves into a self-bound droplet state, in principle there is no need to trap the system and the breathing mode can be interpreted as the response of the droplet to an infinitesimal trapping frequency. With this in mind we will use Eq.~(\ref{eq:sum_rule}) around $\omega_{ho}=0$. Below we will 

discuss  two alternative approaches for estimating the breathing mode, starting  with a phenomenological classical approach and pursuing with a generalized Gross-Pitaevskii equation. 
The phenomenological 
approach  allow us to predict analytically the scaling form of the breathing mode frequency at large number of particles, while a more accurate numerical treatment is achieved  using the GGPE.

\subsection{Phenomenological treatment of the breathing mode}\label{sec:phenomenological}

Let us consider a one-dimensional gas with energy per unit length $e(n)$ for a particle density $n$. The gas can form a stable droplet of density $n_1$  such that
\begin{equation}
  e(n_1)-n_1 \frac{de}{d n_1}=0. 
\end{equation}
We assume that the droplet oscillates while keeping its density uniform, {\it i.e.} always remains in the regime of a flat top density, with edges of negligible size so that
\begin{equation}
\label{eq:density-ansatz}
n(x,t)=\frac{N}{L(t)} \Theta\left(\frac {L(t)}2 -|x|\right), 
\end{equation}
where {$\Theta$ is the Heaviside step function}, $L$ is the droplet size, giving by the continuity equation $\partial_t n + \partial_x j =0$, the current
\begin{equation}
  \label{eq:current-ansatz}
  j(x,t)=\frac{N \dot{L}(t)}{L^2(t)} x,  
\end{equation}
and the velocity $v(x,t)=j(x,t)/n(x,t)$
\begin{equation}
  \label{eq:velocity-ansatz}
  v(x,t)=\frac{\dot{L}(t) x}{L(t)}.   
\end{equation}
Then, the kinetic energy of the oscillating droplet is
\begin{eqnarray}
  \label{eq:kinetic-ansatz}
  T&=&\frac 1 2 m \int_{-L(t)/2}^{L(t)/2} n(x,t) v(x,t)^2 dx \\
  &=& \frac{m N \dot{L}(t)^2}{24},   
\end{eqnarray}
while the potential energy is 
\begin{equation}
  \label{eq:potential-ansatz}
  V=L(t) e\left(\frac N {L(t)}\right).  
\end{equation}
So from the Lagrangian $\mathcal{L}=T-V$ we obtain the equations of motion
\begin{equation}
  \frac{Nm}{12} \ddot{L} = \frac{N}{L(t)} e'\left(\frac N {L(t)}\right)-e\left(\frac N {L(t)}\right). 
\end{equation}
For $N/L=n_1$, we recover the time-independent solution. For densities close to $n_1$, we linearize the equation writing $n(x,t)=\frac{N}{L_{eq}+\delta L(t)} = n_1 (1-\delta L(t)/L_{eq})$ and
\begin{equation}
   \frac{Nm}{12} \delta \ddot{L} = -n_1^3 e''(\rho_1) \frac{\delta L(t)}{N}.  
 \end{equation}
 From the above equation, we obtain the breathing mode frequency:
 \begin{equation}
   \Omega_b^2 = \left(\frac{12 n_1^3 e''(n_1)}{m N^2}  \right), 
   \label{eq:breathing_mode_hydro}
 \end{equation}
 so that $\Omega_b \propto 1/N$. When the profile density shows the flat top behavior, increasing the number of particles corresponds only to a variation of the volume (enlargement of the droplet, (see panel $a)$ of Fig.~\ref{fig:n0_evol}), and therefore in the above estimate of the breathing mode, only the number of particles changes.
 
 In the presence of a weak harmonic trapping we need to add a term to the energy given by,
 \begin{equation}
 U=\frac 1 2 m \omega_{ho}^2 \int dx  x^2 n(x,t), 
 \end{equation}
 and using the above ansatz~(\ref{eq:density-ansatz}) it results in 
 \begin{equation}
U=N \frac{m\omega_{ho}^2}{24} L^2(t).   
 \end{equation}
The ansatz for the density remains valid as long as the harmonic trap does not deform too strongly the droplet, that is as long as $U\ll V$. The modified equilibrium density is then found by minimizing $U+V$ with respect to $L$. 
To first order, we have 
\begin{equation}
n_{eq}(\omega_{ho})=\frac{N}{L_{eq}(\omega_{ho})}=n_1 +\frac{m (\omega_{ho} N)^2}{12n_1^2 e''(n_1)},
\end{equation}
and the condition of validity of our ansatz becomes $\omega_{ho} \ll \Omega_b$. 
With the new Lagrangian $\mathcal{L}=T-U-V$, we obtain
\begin{equation}
\Omega^2(\omega_{ho})=\Omega_b^2 + \omega_{ho}^2\left(4+\frac{n_1 e'''(n_1)}{e''(n_1)}\right).    
\label{eq:breathing_mode_trap_ph}
\end{equation}
Given that $\Omega_b^2 = O(N^{-2})$ in the absence of the trap, Eq.~(\ref{eq:breathing_mode_trap_ph}) is only applicable when 
\begin{equation}
N^2 \ll \frac{12n_1^3 e''(n_1)}{m\omega_{ho}^2}.
\end{equation}
In the Appendix \ref{app:deformation} we briefly sketch
how to take into account the deformation of the droplet by the
harmonic trapping.

 \subsection{Generalized Gross-Pitaevskii approach} 
\label{sec:ggpe}
The previous approximation assumes a discontinuity of the density at the edges of the droplet (flat top)
and neglects entirely the contribution of the density gradients to the potential energy. 
In the present subsection, we introduce a generalized Gross-Pitaevskii equation\cite{kolomeisky2000,dunjko_bosons1d,ohberg_dynamical_2002,oldziejewski_strongly_2019} that allows to take into account the contribution of density gradients to the breathing frequency. 
 The generalized Gross-Pitaevskii equation is obtained by minimizing the action
 \begin{eqnarray}\label{eq:action_gpe} 
 S=\int dx dt \left[i\hbar \psi^* \partial_t \psi  -\frac{\hbar^2}{2m} |\partial_x \psi|^2 -e(|\psi|^2) \right], 
 \end{eqnarray}
 with respect to infinitesimal variation of the field $\psi(x,t)$, where $e(n)$ is the same energy per unit length as in the previous subsection. Using the density phase representation
 \begin{equation}
     \psi(x,t)=\sqrt{n(x,t)} e^{i\theta(x,t)}, 
 \end{equation}
 the variational equations are recast in the hydrodynamic form
 \begin{eqnarray}
 \label{eq:continuity} 
&& \partial_t n +\frac{\hbar}{m} \partial_x (n \partial_x\theta)=0 \\
\label{eq:bernouilli}
&& \frac{\hbar} m \partial_t \theta +\frac 1 2 \left(\frac{\hbar} m \partial_x \theta\right)^2 + \frac{e'(n)}{m} -\frac{\hbar^2}{2m^2 \sqrt{n}} \partial_x^2(\sqrt{n}) =0 \nonumber \\ 
 \end{eqnarray}
 The equilibrium solution is found with $\theta(x,t)=-\mu t/\hbar$. 
 Now, if $n_0(x)$ is a variational solution describing a static droplet, we can {assume} a time dependent density of the form
 \begin{equation}\label{eq:density} 
     n(x,t)=\frac 1 {R(t)} n_0\left(\frac x {R(t)} \right),  
 \end{equation}
{with $R(t)$ dimensionless,} that describes a droplet that is stretched or compressed as a function of time, containing $N=\int dx n_0(x)$ particles.  
 The continuity equation (\ref{eq:continuity}) then imposes a space and time dependent phase 
 \begin{equation}
     \theta(x,t)=\frac{m x^2}{2\hbar} \frac{\dot{R}(t)}{R(t)} + \int_0^t \frac{\mu(t') dt'}{\hbar}.  
 \end{equation}
 If that ansatz is inserted in the action, Eq.~(\ref{eq:action_gpe}), one finds
 \begin{eqnarray}
 && S=\frac {Nm\langle x^2\rangle} 2 \int dt \dot{R}^2(t) -\int \frac{\hbar^2}{8m R^2(t)} \int dy \frac{n_0'(y)^2}{n_0(y)} \nonumber \\ 
 &&-\int dt R(t) \int dy e\left(\frac{n_0(y)}{R(t)}\right) -N \int dt \mu(t), 
 \end{eqnarray}
 where 
 \begin{equation}
     \langle x^2 \rangle = \frac{\int dy y^2 n_0(y)}{\int dy n_0(y)}. 
 \end{equation}
 For $R(t)=1$, and $d\mu/dt=0$, the action is already extremal since $n_0(x)$ is a static solution. 
 By extremizing $S$ with respect to $R(t)$, one has the equation of motion 
 \begin{eqnarray}\label{eq:eom_varia} 
 && N m \langle x^2 \rangle \ddot{R}(t)=\frac{\hbar^2}{4m R^3(t)} \int dy \frac{n_0'(y)^2}{n_0(y)} \nonumber \\ && - \int dy \left[e\left(\frac{n_0(y)}{R(t)}\right) -\frac{n_0(y)}{R(t)} e'\left(\frac{n_0(y)}{R(t)}\right)\right]. 
 \end{eqnarray}
 In the Appendix~\ref{app:ggpe-gs}, we check that when $n_0(y)$ is the ground state density, Eq.~(\ref{eq:eom_varia}) is satisfied with $R=1$. 
 In the case of small oscillations of the droplet, $R(t)=1+\epsilon \cos(\Omega_b t)$, with $\epsilon \ll 1$, we are lead to the breathing frequency
 \begin{eqnarray}
 \Omega_b^2 = \frac{\int dy n_0(y)^2 e"(n_0(y)) + \frac{3\hbar^2}{4m} \int \frac{dy}{n_0(y)}\left(\frac{dn_0(y)}{dy}\right)^2}{m \int dy y^2 n_0(y)}. 
 \label{eq:breathing_mode}
 \end{eqnarray}
 In Eq.~(\ref{eq:breathing_mode}), only the equilibrium density profile enters, allowing to compute the frequency of the breathing mode from the minimization of the static generalized Gross-Pitaevskii Hamiltonian. 
The result of the phenomenological treatment of sec.\ref{sec:phenomenological} is recovered when we assume $n_0(x)$ constant inside the droplet (flat top) and zero outside. When the profile differs significantly from a flat top with narrow edges, we can expect from Eq.~(\ref{eq:breathing_mode}) deviations from the phenomenological Eq.~(\ref{eq:breathing_mode_hydro}). In particular, this should be the case at low number of particles with a so-called 
{bright soliton-like} profile. 
The variational treatment can be easily generalized to
the presence of a harmonic trap with frequency $\omega_{ho}$. By adding a term $- V_{ext} (x)|\psi(x)|^2$ to the action Eq. (12) and
using the density-phase representation, one arrives to the
result:
\begin{equation}
\label{eq:breathing-in-trap} 
\Omega^2(\omega_{ho})=\omega_{ho}^2 + \frac{\int dy n_0(y)^2 e"(n_0(y)) + \frac{3\hbar^2}{4m} \int \frac{dy}{n_0(y)}\left(\frac{dn_0(y)}{dy}\right)^2}{m \int dy y^2 n_0(y)},
\end{equation}
where the profile density $n_0(y)$, and $\Omega_b$ consequently, is calculated using the generalized Gross-Pitaevskii equation including the trapping potential. A fully quantum treatment using the Dirac-Frenkel variational principle\cite{dirac_variational1930,frenkel_book1950} with a many-body trial wavefunction obtained by applying a dilatation to the ground state wavefunction is discussed in App.~\ref{app:tdva}. The generalized Gross-Pitaevskii equation can be understood as the classical limit of that approach. In the fully quantum treatment, the breathing mode can be linked to higher moments $m_3/m_1$ of the dynamic structure factor, as shown in App.~\ref{app:tdva} 

\section{Results}
\label{sec:results}
All derivations in the previous Section assume the formation of droplets with a flat top density profile and are extended for situations where the latter is not altered by the presence of a trapping potential. 
Here we test the predictions and the expressions for the breathing mode by using the ground-state energy based on the Variational Bethe-Ansatz\cite{depalo2020,depalo2021,depalo2022,depalo2023} to calculate the density profile and plug it in the sum rule  Eq.~(\ref{eq:sum_rule}). 
We choose $\mu_D = 9.93 \mu_B$, $l_\perp = 57.3nm$, {$\omega_\perp=2\pi 19 kHz$} and $a_{ho} =\sqrt{\hbar/(m \omega_{ho})} = 24000 a_0$, where $a_0$ is the Bohr radius, to make contact with recent experimental works on the strictly one-dimensional dipolar $\mathrm{{}^{162}Dy}$ gas\cite{Lev2020,Lev_therm_2018}. Finally we consider the case in which the angle between the dipoles orientation and the longitudinal z-axis is $\theta=0$.

\subsection{Breathing mode without a trapping potential}
\label{sec:breath-notrap}
\begin{figure}[h]
\begin{center}
\includegraphics[width=85.mm]{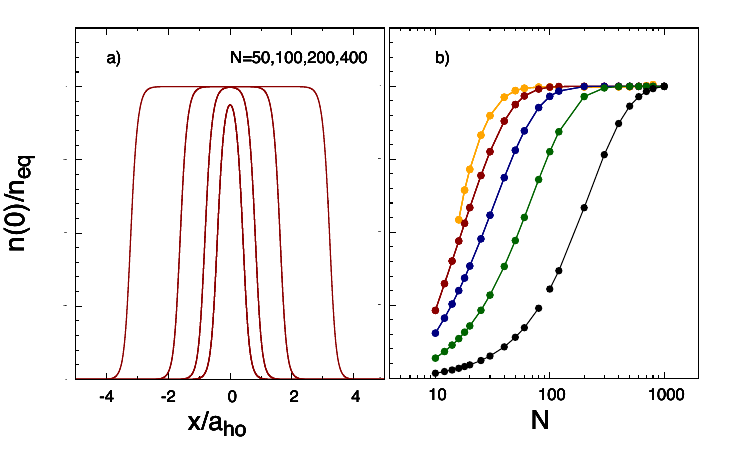}
\end{center}
\caption{ 
In panel $a)$ we show the evolution of the density profile  as the number of the particles in the cloud is increased for the case 
$a_{1D}=-7000 a_0$, for some selected cases {\it i.e.} $N=50,100,200$ and $400$. In panel $b)$ we show, for several scattering lengths $a_{1d}$, the ratio between the density at the center $n(x=0)$ of the cloud and the equilibrium density reached when the density profile becomes flat as a function of the numbers of particle $N$ in the droplet. Data for scattering lengths $a_{1d}/a_0=-6500,-7000,-7500,-8000,-8500$ are shown using 
orange, dark-red, dark-blue, dark-green and black solid dots respectively.
All lines are intended to be guide to the eye.}
\label{fig:n0_evol}
\end{figure}

We follow the formation of the flat-top droplets by looking at the density profiles obtained after solving the stationary generalized Gross-Pitaevskii equation, generalized by replacing the Hartree-term with the energy per unit length of the bulk quasi-one dimensional dipolar system obtained using  the Variational Bethe Ansatz approach\cite{depalo2020,depalo2022}.
In panel a) of Fig.~\ref{fig:n0_evol} we follow the evolution of the density profile as a function of the number of particles in the cloud, namely $N=50,100,200$ and $400$, for the case $a_{1D}=-7000 a_0$, in the absence of trap. The system supports the occurrence of a self-bound droplet ($\mu < 0$) and, on increasing the number of particles $N$, the typical flat-top shape is reached ($\mu < 0$ and constant). 
In the left panel $b)$ we display this behavior for different scattering lengths, by showing the value of the density at the center of the trap, normalized by the equilibrium density reached when the density profile becomes flat, as a function of $N$. The flat-top density profile is reached when the density at center of the cloud is constant.

\begin{figure}[h]
\begin{center}
\includegraphics[width=85.mm]{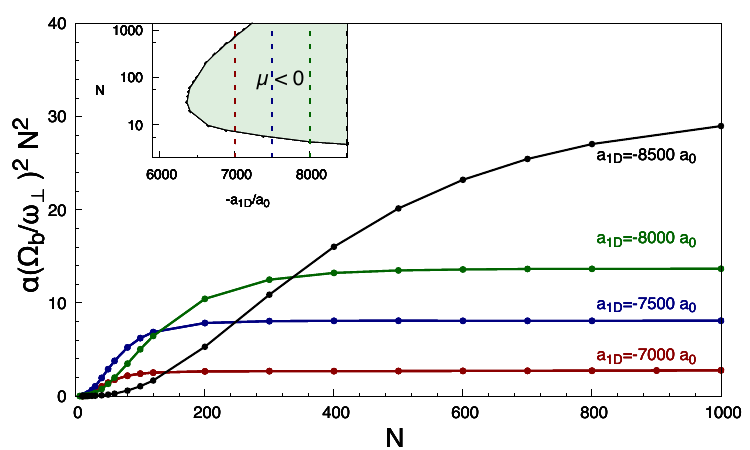}
\end{center}
\caption{Breathing mode square of the droplet in the absence of trapping potential multiplied by $N^2$ for several value of $a_{1D}$. Dark-red, blue, dark-green and black solid dots are estimates of {$\Omega_b^2/\omega_\perp$} using sum-rule approach from Eq.~(\ref{eq:sum_rule})
for $-a_{1D}/a_0=7000,7500,8000,8500$, { using as multiplicative factors,
$\alpha=1,1/3,1/30,1/900$ respectively.}
In the inset we show the phase diagram for the dipolar system from Ref.~\onlinecite{depalo2022}, 
where the self-bound droplet occurs when the chemical potential $\mu < 0$.
The dashed lines corresponds to the values shown in the main figure with the same color code. All lines are intended to be guide to the eye.}
\label{fig:flat_wb_n}
\end{figure}

We verify the $N^{-1}$ scaling hydrodynamic prediction for the 
breathing mode, showing that $\Omega^2_b N^2$ estimated using 
sum-rule Eq.~\ref{eq:sum_rule} saturates to a constant value for
large number of particles, in the droplet phase (see
Fig.~\ref{fig:flat_wb_n}).

\begin{figure}[h]
\begin{center}
\includegraphics[width=85.mm]{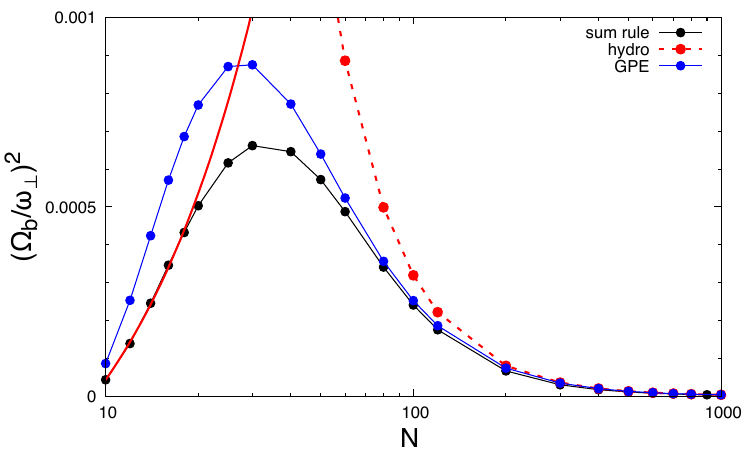}
\end{center}
\caption{Breathing mode of the droplet as a function of number 
of particles $N$ without the trap for $a_{1D}=-7000 a_0$. 
{($\omega_\perp$ units)}
Black, blue and red solid dots are estimates for $\Omega_b$ 
using the sum-rule approach Eq. ~(\ref{eq:sum_rule}), derivation 
from Gross-Pitaevskii  Eq.~(\ref{eq:breathing_mode}) and the
hydrodynamic approach Eq.~(\ref{eq:breathing_mode_hydro})
respectively. {Solid red line is a fitting curve $\Omega_b(N)=a (N-N_0)^\beta $ with $\beta=0.49(3)$.}
All the lines between points are intended to be guide to the eye.}
\label{fig:wb_cfr_7000}
\end{figure}

We see in more detail the behavior of the breathing mode comparing its estimates 
obtained using the different approximations for a selected scattering length, namely
$a_{1D}=-7000 a_0$, as a function of the number of particles {(see Fig.~\ref{fig:wb_cfr_7000})}. For $N \gtrsim 100$
we can identify a region, consistent with a flat-top density profile region (see
panel $a)$ of Fig.\ref{fig:n0_evol}) where all the three estimates used collapse.
The fact that all the approaches quantitatively agrees with each other implies that the response function
$\mathrm{Im}\chi(\omega) \sim Z \delta(\omega-\Omega_b)$, see App. \ref{app:holder},
and {\it i.e.} in this region the breathing mode is sharply
defined. 

When the number of particles in the cloud is smaller, yet larger than a critical 
value beyond 
which the  system is not self-bound (in this case roughly $N\simeq 10$ see inset of
Fig.~\ref{fig:flat_wb_n}), the profile density looses the flat-top droplet shape. The 
phenomenological hydrodynamic approach strongly relies on the assumption that the droplet
oscillates while keeping its density uniform (i.e. the droplet shape is a simple rectangle).
In the small $N$ region, where the cloud has not achieved a flat-top density profile, the
breathing mode grows as $N^\beta$ with {$\beta < 2/3 $} where $\beta=2/3$ is the behavior 
predicted for a solitonic shape density profile\cite{Malomed_2018}. 
For large number of particles where the droplet shows a flat top, the $N^{-1}$ decay 
of the breathing mode predicted in Eq.~(\ref{eq:breathing_mode_hydro}) works well but
becomes obviously inaccurate when we depart from the simple rectangular shape and it
is necessary to retain the whole profile density as in the generalized Gross
-Pitaevskii equations (Eq.~\ref{eq:breathing_mode}). The two 
different behaviors of the breathing mode as a function of $N$
produce a maximum that can be interpreted, together with the $N^{-1}$
scaling, as the signature of the development of a droplet\cite{Malomed_2018,Malomed_2020,Xucong_2023}. 

In Fig.~\ref{fig:wb_cfr_7000} we see that the estimate of the breathing mode based on Eq.~(\ref{eq:breathing_mode}) is always above the sum-rule estimate Eq.~(\ref{eq:sum_rule}) in the whole $N$ range; we expect this hierarchy since the generalized Gross-Pitaevskii estimate is related to the moment ratio $m_3/m_1$ while the sum-rule is related to the ratio $m_1/m_{-1}$, and $m_3/m_1 \ge m_1/m_{-1}$ as discussed in App.~\ref{app:holder}. 

\begin{figure}[h]
\begin{center}
\includegraphics[width=95.mm]{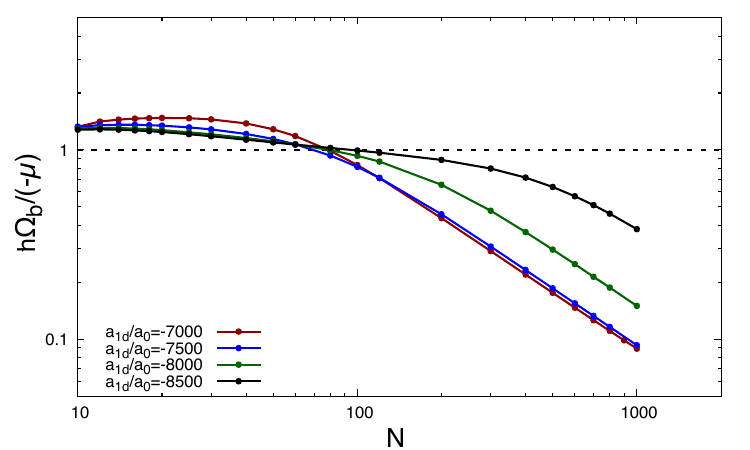}
\end{center}
\caption{Ratio of breathing mode, using the sum-rule approach, of the droplet and chemical potential as a function of number of particles $N$ without the trap for 
$-a_{1D}/a_0=7000,7500$ and $8500$ ( Dark-red, blue, dark-green and black solid dots respectively). All lines are intended to be guide to the eye.}
\label{fig:notrap_wb_mu_sr} 
\end{figure}

We conclude this part of the section by commenting on the stability of the cloud 
and therefore on the possibility of detecting the droplet and this behavior for an experimental configuration close to the one achieved in Ref.~\onlinecite{Lev_therm_2018,Lev2020}. We look at the ratio of 
the breathing-mode frequency to the particle-emission threshold:
$-\hbar \Omega_b/\mu$ which is shown in  Fig.~\ref{fig:notrap_wb_mu_sr} for several scattering lengths as a function of the number of particles in the cloud. 
When the energies of the collective mode are above the chemical potential, {\it i.e.} $-\hbar \Omega_b/\mu > 1$, exciting the droplet leads to emission of particles. As a consequence the cloud self-evaporates \cite{petrov2015}, a property that might make the system a possible bath for sympathetic cooling for the other system.
In a dipolar bosonic gas confined in a cigar-shaped trap\cite{Baille2017}, or in a strictly 1D bosonic mixture\cite{Malomed_2020} the breathing mode was  predicted to be always bound.

By contrast, in the case we are investigating, which is strictly one-dimensional and where besides the dipolar interaction we also have the repulsive contact interaction, we find that for size smaller than $N \simeq 80$ exciting the droplet would imply an evaporation of the cloud itself, that could occur by the fragmentation of the droplet into small pieces. 
Therefore the increasing scaling of the breathing frequency as a function of $N$, as well as its maximum will not   be detectable, while for larger number of particles, yet achieveable, the cloud is stable and this occurs when the flat-top density profile is reached and  the $N^{-1}$ behavior of the breathing mode fullfilled.

\subsection{ Breathing mode for a trapped system}\label{sec:breath-trap}

In this part of the section we analyze how the trapping potential affects the density profile and consequently the breathing mode and if it is possible to detect a scaling law similar to the one obtained for the self-bound droplet.
As discussed in the previous section the first effect of the trapping potential is to introduce $\omega_{ho}$ in the breathing potential
as the relevant scale as shown in Eq.~(\ref{eq:breathing_mode_trap_ph}) and Eq.~(\ref{eq:breathing-in-trap}).

\begin{figure}[ht]
\begin{center}
\includegraphics[width=90.mm]{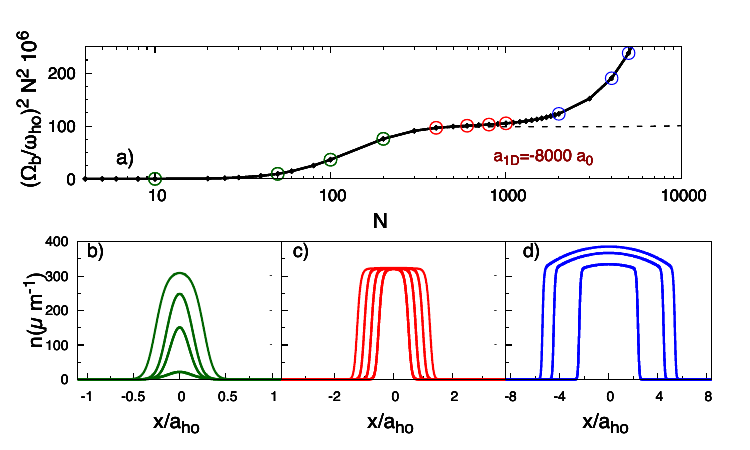}
\end{center}
\caption{ In the upper panel $a)$ we show $ \alpha(\Omega_b/\omega_{ho})^2 N^2$ for the scattering length $a_{1D}=-8000 a_0$. {The dashed black curve is the same quantity computed in the absence of longitudinal trapping, exhibiting a plateau for large $N$.
In the lower panels we show the density profiles of the trapped cloud for different number of particles in different regions.  Density profiles are represented by solid dark-green lines for $N=10,50,100$ and $200$ (panel $b)$, by red solid lines for $N=400,600,800$ and $1000$ (panel $c)$ and by blue solid lines for $N=2000,4000$ and $5000$ (panel $d)$. Open dots in the upper panel correspond to results for the same number of particles as in the density profile shown in the lower panels with identical color code.
Lines are intended to be guide to the eye.}}
\label{fig:wb_trap_8000}
\end{figure}

In Fig.~\ref{fig:wb_trap_8000} we show a case where there is an appreciable
region where the effect of the potential is not so strong and we can 
appreciate a plateau in the quantity $\Omega_b^2 N^2$ that {corresponds} to 
the region of $N$ for which we observe 
a flat top behavior in the density profile (see red data in panels $a)$ and
$c)$). When $N \gtrsim 1000 $ the effect of the trap {on} the density profile is
visible (see panel $d)$), the chemical potential becomes eventually positive,
and we {lose} the plateau in $\Omega_b^2 N^2$.

The qualitative picture of the breathing mode as a function of the number of particles is
similar to the one obtained in the absence of the trapping potential: there is plateau in 
$\Omega_b N$ that signals the fact that droplet with a flat-top density profile is developing.

We conclude our analysis by comparing the different approaches for computing the breathing mode for trapped system using, beyond the 
standard sum-rule approach, Eq.~(\ref{eq:sum_rule}), the hydrodynamic expression 
Eq.~(\ref{eq:breathing_mode_trap_ph}) and the generalized Gross-Pitaevskii 
approach Eq.~(\ref{eq:breathing-in-trap}).
Analogously to the case in the absence of the trap, we observe that breathing-mode estimated by using the sum rule has the lowest value and hierarchy between the estimates of $\Omega_b$ is fulfilled as well (App.~\ref{app:holder}).
\begin{figure}[h]
\begin{center}
\includegraphics[width=85.mm]{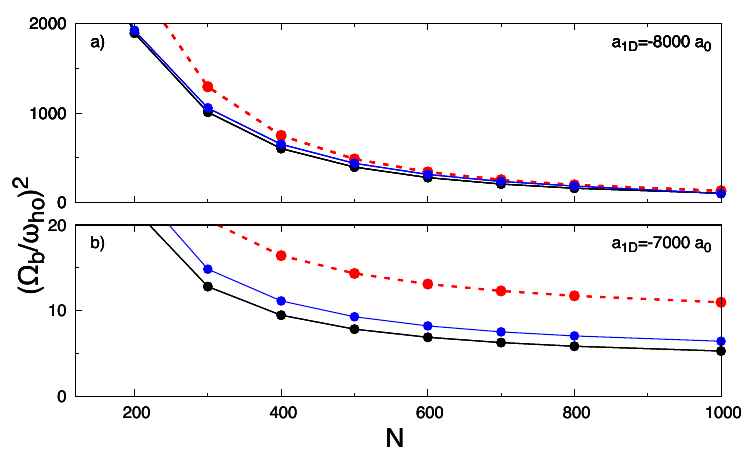}
\end{center}
\caption{ $\Omega^2_b/\omega^2_{ho}$ as a function of number 
of particles $N$ with the trap for $a_{1D}=-8000 a_0$ (panel $a)$)
and $a_{1D}=-7000 a_0$ (panel $b)$).
Black, blue and red solid dots are estimates for $\Omega_b$ 
using the sum-rule approach Eq.~(\ref{eq:sum_rule}), derivation from 
Gross-Pitaevskii  Eq.~(\ref{eq:breathing-in-trap}) and the hydrodynamic approach Eq.~(\ref{eq:breathing_mode_hydro}) respectively. All lines are intended to be guide to the eye.}
\label{fig:wb_cfr_tails_trap}
\end{figure}
As largely already discussed before, in the absence of trap the breathing mode of the droplets decays as $N^{-1}$ and all the different estimates converge 
(see fig.~\ref{fig:wb_cfr_7000}) once we are in the region where the droplet has a flat-top density profile. 
We can observe the same kind of collapse only if we are in the flat-top region also with the trapping potential, that means that the effect of the trap is small and therefore Eq.~(\ref{eq:breathing_mode_hydro}) and Eq.~(\ref{eq:breathing-in-trap}) are good approximations and the collective mode is well defined. 
This is clearly shown in Fig.~\ref{fig:wb_cfr_tails_trap}, where in panel $a)$ we show $(\Omega_b/\omega_{ho})^2 $ for $a_{1D}/a_0=8000 $ in a range of $N$ where 
the flat-top shape of the profile density holds (see Fig.~\ref{fig:wb_trap_8000}) with the appropriate consequences. 
In this range there in an apparent collapse. However, in the same range of $N$ we can consider a scattering length smaller in absolute value ($a_{1D}/a_0=-7000$). In this case the trapping potential affects the profile density shape so we cannot expect the collapse of all derivations (see panel $b)$ of Fig.~\ref{fig:wb_cfr_tails_trap}. 

\section{Conclusions}
\label{sec:conclusions}
We have investigated the breathing mode of a quantum droplet in a one-dimensional dipolar Bose gas in a regime  close to that experimentally realized\cite{Lev_therm_2018,Lev2020}. Our theory relies on a variational calculation of the ground state energy as a function of density\cite{depalo2020,depalo2021,depalo2023}. When the dipolar interactions are made attractive, by increasing the scattering length and the number of particles, a soliton-like density profile is obtained with a maximum  that scales linearly with the number of particles $N$. Upon further increasing the number of particles, the density acquires  a flat-top profile  whose maximum does not vary with $N$, indicating the formation of the droplet. 
Starting with a phenomenological approach, that relies on an idealized rectangular  density profile of the droplet, and simple hydrodynamics,  we have found that the breathing mode frequency  scales as $N^{-1}$. We have then turned to a generalized Gross-Pitaevskii (GGPE) approach\cite{kolomeisky2000,oldziejewski_strongly_2019} that takes into account explicitly density gradients. Finally we have benchmarked the results with the sum-rule approach, in the limit of zero trapping confinement. Of these three methods, the sum rule always gives the lowest estimate, as it links the breathing mode to the lowest moments of the dynamic structure factor. Both the sum-rule and the GGPE give a vanishing breathing frequency at low particle number $N$ and show a maximum of the breathing frequency at intermediate values of $N$. At large value of $N$, they become asymptotic to the hydrodynamic approximation. We have considered the stability of the cloud against particle emission, and we have found that the cloud is stable at $N$ sufficiently large.  We have also analyzed the breathing mode of the droplet in the presence of longitudinal trapping. In the presence of the trap the frequency of the breathing mode does not return to zero when increasing the number of particles, saturating to a value that depends on the scattering length.  One remark is in order here: While the {density profile} is the natural observable to investigate the formation of droplets in two or three dimensions where the breaking of the translational symmetry is well established, the quasi-one dimensional case is more delicate because the droplet corresponds to a flat top density and the measure of the breathing mode we propose is more reliable. We believe that our theoretical predictions for quasi-one-dimensional quantum droplets are testable with the available experimental setups of strongly confined gases of dipolar atoms in quasi-one dimensional geometry. In particular, the behavior of the measured breathing mode of the droplet will highlight the reliability and accuracy of the three theoretical methods we have proposed. Finally, we stress that the generalized Gross-Pitaevskii approach developed here could be applied also to droplets forming in other quasi-one-dimensional superfluid systems.

\section*{Acknowledgements} 

LS and RC are partially supported by “Iniziativa Specifica Quantum” of INFN. LS is partially supported by the Project "Frontiere Quantistiche" (Dipartimenti di Eccellenza) of the Italian Ministry for Universities and Research, by the European Quantum Flagship Project "PASQuanS2", by the European Union-NextGenerationEU within the National Center for HPC, Big Data and Quantum Computing [Spoke 10: Quantum Computing], by the BIRD Project "Ultracold atoms in curved geometries" of the University of Padova, and by PRIN Project "Quantum Atomic Mixtures: Droplets, Topological Structures, and Vortices". RC and SDP acknowledge the Institute Henri Poincaré for hospitality during the completion of this work.  

\appendix
\section{Dirac-Frenkel variational method and sum rules}\label{app:tdva} 
In a second quantized treatment, we can introduce the dilatation operator 
\begin{equation}\label{eq:dilatation} 
  D=\frac{\hbar}{2i} \int dx x \left(\psi^\dagger \frac{\partial\psi}{\partial x} - \frac{\partial\psi^\dagger}{\partial x}\psi \right),  
\end{equation}
and the quadrupole operator 
\begin{equation}\label{eq:quadrupole} 
  Q=\frac 1 2 \int dx x^2 \psi^\dagger(x) \psi(x),  
\end{equation}
satisfying the commutation relations 
\begin{eqnarray}
\label{eq:quadham} 
  [Q,H] &=&\frac{i \hbar}{m} D, \\
  \label{eq:quaddil} 
  [D,Q]&=&-2i \hbar Q.     
\end{eqnarray}
If we consider the annihilation operator $\psi(x)$, the  particle density $n(x)$ and the particle current $j(x)$, 
they transform under a time-dependent dilatation as 
\begin{eqnarray}
\label{eq:dilannh}
  e^{i\frac{\lambda D}{\hbar}} \psi(x) e^{i \frac{m Q}\hbar \frac{d\lambda}{dt}} e^{-i \frac{m Q}\hbar \frac{d\lambda}{dt}} e^{-i\frac{\lambda D}\hbar} &=& e^{-i \frac{m x^2}{2\hbar} \frac{d\lambda}{dt}} e^{-\lambda/2} \psi(x e^{-\lambda}), \\    
\label{eq:dildens}
  e^{i\frac{\lambda D}{\hbar}}   e^{-i \frac{m Q}\hbar \frac{d\lambda}{dt}}  n(x) e^{i \frac{m Q}\hbar \frac{d\lambda}{dt}} e^{-i\frac{\lambda D}\hbar} &=&e^{-\lambda} n(xe^{-\lambda}), \\ 
  \label{eq:dilcur}
   e^{i\frac{\lambda D}{\hbar}}   e^{-i \frac{m Q}\hbar \frac{d\lambda}{dt}}  j(x) e^{i \frac{m Q}\hbar \frac{d\lambda}{dt}} e^{-i\frac{\lambda D}\hbar} &=& e^{-2\lambda} j(x e^{-\lambda}) \\ && + \frac d {dt} (e^{-\lambda}) x n(x e^{-\lambda})\nonumber .  
\end{eqnarray}
Choosing as normalized trial wavefunction 
\begin{eqnarray}\label{eq:trial-wf} 
  |\Psi\rangle(t)=e^{\frac{imQ}{\hbar} \frac{d\lambda}{dt} } e^{-i \lambda(t) D/\hbar} e^{-i E_0 t/\hbar} |0\rangle,  
\end{eqnarray}
the expectation values of the particle density and current behave as 
\begin{eqnarray}
  \langle \Psi |\rho(x) |\Psi\rangle &=& e^{-\lambda} \langle 0|\rho(x e^{-\lambda})|0\rangle = \frac 1 {R} n_0\left(\frac x R \right),\nonumber \\ 
 \langle \Psi |j (x) |\Psi\rangle &=& e^{-2\lambda} \langle 0|j(x e^{-\lambda})|0\rangle+ \frac d {dt} (e^{-\lambda}) x  \langle 0|\rho(x e^{-\lambda}) |0\rangle \nonumber \\
&=&0 + \frac{mx \dot{R}}{\hbar R^2 } n_0\left(\frac x R \right),
\end{eqnarray}
so we recover the ansatz we used for the density in the GGPE with $R(t)=e^{\lambda(t)}$. Moreover, using the representation
\begin{equation}\label{eq:densphas} 
    \psi(x)=e^{i\theta(x)} \sqrt{n(x)}
\end{equation}
with $[\theta(x),\rho(x')]=i \delta (x-x')$ the time dependent rescaling shifts the phase $\theta$ by $m x^2 \dot{\lambda}(t)/(2\hbar)$ as in the GGPE. 
Inserting the trial wavefunction~(\ref{eq:trial-wf}) in the Dirac-Frenkel\cite{dirac_variational1930,frenkel_book1950} action 
\begin{eqnarray}\label{eq:dirac-frenkel}
  S=\int dt \left[i \hbar \langle \Psi | \partial_t   |\Psi\rangle -  \langle \Psi | H |\Psi\rangle \right],  
\end{eqnarray}
we get 
\begin{eqnarray}
&& S=  m \langle 0 | Q | 0\rangle \int dt  \left[\frac{d}{dt}(e^\lambda)\right]^2 \nonumber \\
&& - \int dt  \langle 0|  e^{\frac i \hbar \lambda D} (H-E_0) e^{-\frac i \hbar \lambda D} |0\rangle,  \nonumber
\end{eqnarray}
and for small oscillations, 
\begin{eqnarray}
  S=m \langle 0 | Q | 0\rangle \int dt  \left(\frac{d\lambda}{dt}\right)^2 -   \int dt \frac{\lambda^2}{\hbar^2} \langle 0| D (H-E_0) D |0\rangle,\nonumber 
\end{eqnarray}
yielding
\begin{eqnarray}\label{eq:breathing-q}
\Omega_b^2 &=& \frac{\langle 0| D (H-E_0) D |0 \rangle}{\hbar^2 \langle 0|Q|0\rangle}, \nonumber \\ 
&=& \frac{\langle 0 | Q (H-E_0)^3 Q |0\rangle}{\langle 0 | Q (H-E_0) Q |0 \rangle}, 
\end{eqnarray}
 where in the second line, we have used the commutation relations~(\ref{eq:quadham})--(\ref{eq:quaddil}). That last expression can be rewritten as a sum-rule\cite{stringari_sumrules}. Using the imaginary part of the quadrupole-quadrupole response function, 
 \begin{eqnarray}
  \mathrm{Im}\chi_{QQ}(\omega) = \pi \sum_n |\langle 0 | Q |n\rangle |^2 \delta (E_n -\hbar \omega -E_0), 
\end{eqnarray}
we rewrite our expression 
\begin{eqnarray}\label{eq:sumrule} 
\Omega_b^2 = \frac{\int_0^{+\infty}  d\omega \omega^3 
\mathrm{Im}\chi_{QQ}(\omega)}{\int_0^{+\infty}  d\omega \omega  \mathrm{Im}\chi_{QQ}(\omega)}.   
\end{eqnarray}

\section{Sum rules and H\"{o}lder inequality}\label{app:holder}
Sum rules such as~(\ref{eq:sumrule}) or the expressions derived in Ref.~\onlinecite{stringari_sumrules} yield  upper bounds for the gap between the ground state and  the first excited state having a nonzero matrix element of $Q$ with the ground state 
\begin{eqnarray}\label{eq:upperbound} 
  E_1-E_0 \le \Delta(\alpha,\beta) =\left[ \frac{\int_0^{+\infty} d\omega (\hbar \omega)^\alpha  \mathrm{Im}\chi_{QQ}(\omega)}{\int_0^{+\infty} d\omega (\hbar \omega)^\beta  \mathrm{Im}\chi_{QQ}(\omega)}\right]^{\frac 1{\alpha -\beta}}.   
\end{eqnarray}
These bounds are related to each other by 
\begin{eqnarray}\label{eq:relbound} 
  \Delta(\alpha,\beta)^{\alpha-\beta} \Delta(\beta,\gamma)^{\beta-\gamma} =\Delta(\alpha,\gamma)^{\alpha-\gamma}, 
\end{eqnarray}
and in the case of a single mode  {$\mathrm{Im} \chi(\omega)= A \delta(\omega-\Omega)$}, they all lead to $E_1-E_0 \le \Omega$. With a richer spectrum, the sum rules result in different estimates for the energy of the first excited state. The bounds Eq.~(\ref{eq:relbound}) depend of the generalized moments 
\begin{equation}
m_\alpha=\int_0^{+\infty} d\omega (\hbar \omega)^\alpha  \mathrm{Im}\chi_{QQ}(\omega)
\end{equation}
of the response function. For $\alpha$ a positive integer, the moment $m_\alpha$ can be expressed using only  multiple commutators of $Q$ with the Hamiltonian, while for $\alpha=-1$ it can be expressed in terms of the real part of the response function $\chi_{QQ}(\omega=0)$. This makes such bounds convenient since they only depend on quantities computed in the ground state. However, one would like to determine which bounds yield the most accurate estimates for the energy of the excited state. It turns out that using H\"older inequalities\cite{abramowitz_math_functions}, such hierarchy can be established. 
If we start from the integral 
\begin{eqnarray}
   \int_0^{+\infty}  \frac{d\omega}{\pi} (\hbar \omega)^ {x\alpha+(1-x)\gamma} \mathrm{Im}\chi(\omega),  
\end{eqnarray}
with $0<x<1$
by choosing $p=1/x$ and $q=1/(1-x)$, and applying the  H\"older inequalities (3.2.8) or (3.2.10) of Ref.\cite{abramowitz_math_functions}  to the functions
\begin{eqnarray}
  (\hbar \omega)^{\alpha/p} \left(\mathrm{Im}\chi(\omega)/\pi\right)^{1/p}, 
\end{eqnarray}
and
\begin{eqnarray}
  (\hbar \omega)^{\gamma/q} \left(\mathrm{Im}\chi(\omega)/\pi\right)^{1/q}, 
\end{eqnarray}
we find
\begin{eqnarray}
   && \int_0^{+\infty}  \frac{d\omega}{\pi} (\hbar \omega)^ {x\alpha+(1-x)\gamma} \mathrm{Im}\chi(\omega) \le   \left[\int_0^{+\infty}  \frac{d\omega}{\pi} (\hbar \omega)^ {\alpha} \mathrm{Im}\chi(\omega) \right]^x \nonumber \\ && \times \left[\int_0^{+\infty}  \frac{d\omega}{\pi} (\hbar \omega)^ {\gamma} \mathrm{Im}\chi(\omega) \right]^{(1-x)}.  
\end{eqnarray}
Such inequality, inserted in Eq.~(\ref{eq:upperbound}), results in 
\begin{equation}
  \Delta(\alpha,\alpha x +\gamma(1-x)) \ge   \Delta(\alpha x +\gamma(1-x),\gamma).  
\end{equation}
In particular, with $\alpha>\beta>\gamma$, setting $x=(\beta-\gamma)/(\alpha-\gamma)$, we get $\Delta(\alpha,\beta)\ge\Delta(\beta,\gamma)$. Then, by  inserting this inequality in  Eq.~(\ref{eq:relbound}), we obtain $\Delta(\alpha,\beta)\ge \Delta(\alpha,\gamma) \ge \Delta(\beta,\gamma)$, so more accurate estimates are obtained with lower values of the exponents in Eq.~(\ref{eq:upperbound}).  
In particular, the estimate given by Eq.~(\ref{eq:sumrule}), that is $\alpha=3,\beta=1$ in Eq.~(\ref{eq:upperbound}), is always above the one with $\alpha=1,\beta=-1$ from Ref.~\onlinecite{chiofalo_sr_charged_bose,menotti_stringari}. Obviously, when the difference between the two approximations becomes large, it is an indication that the breathing mode is not sharply defined. So, by comparing the results of the two approximations, we can identify the regions of the phase diagram where a well defined breathing mode can be observed.   
\section{Deformation of the droplet}\label{app:deformation}
When the condition $\omega_0 \ll \Omega_b$ is not fulfilled, we must take into account the deformation of the droplet by the harmonic trapping. This can be done using a local density approximation\cite{depalo08_dipolar}. The density in the system at rest is given by
\begin{equation}
    e'(\rho_{\mathrm{eq}}(x))=\mu-V(x), 
\end{equation}
but we have also\cite{depalo2022} $e(\rho)\ge e'(\rho_1)\rho$, and $E-\mu N\ge \int [e'(\rho_1)+V(x)-\mu]\rho(x) dx$. So when $V(x)>\mu-e'(\rho_1)$, we can minimize energy by setting $\rho_{\mathrm{eq}}(x)=0$. We have therefore a discontinuous drop of the density from $\rho_1$ to zero at $x=\pm L/2$, and $V(L/2)=\mu-e'(\rho_1)$. At the center of the trap, the density is given by $e'(\rho_{\mathrm{eq}}(0))=\mu$ and the number of particles is 
\begin{eqnarray}
N&=&\sqrt{\frac{2}{m\omega_0^2}} \left[\rho_1 \sqrt{e'(\rho_{\mathrm{eq}}(0){)}-e'(\rho_1)} \right.\nonumber \\ && \left. +\int_{\rho_1}^{\rho_{\mathrm{eq}}(0)} \sqrt{e'(\rho_{\mathrm{eq}}(0))-e'(\rho)} d\rho \right]
\end{eqnarray}
Hydrodynamic equations\cite{menotti_stringari,depalo08_dipolar}
can be derived from the Hamiltonian 
\begin{eqnarray}
H=\int_{-L/2}^{L/2} \left[\frac{\rho_{\mathrm{eq}}(x)}{2m} (\partial_x\theta)^2 + e''(\rho_{\mathrm{eq}}(x)) \frac{(\delta \rho)^2}{2} \right], 
\end{eqnarray}
with commutation relations $[\theta(x),\delta \rho(y)] = -i \delta(x-y)$. The field $\theta(x,t)$ is a velocity potential, while $\delta\rho(x,t)$ is the deviation of the local density from its equilibrium value. In general, an ansatz of the form Eq.~(\ref{eq:velocity-ansatz}{)} corresponds to $\theta(x,t)=\epsilon(t) x^2/2$. 
The continuity equation still leads to an ansatz compatible with Eq.~(\ref{eq:density-ansatz}) for $\delta\rho(x,t)$, but in general, the Euler equation cannot be satisfied. Nevertheless, we can introduce the Lagrangian
\begin{eqnarray}
    \mathcal{L}_{\mathrm{hydro}}=\int dx \left[\frac{(\partial_t\theta)^2}{2e"(\rho_{\mathrm{eq}}(x))} -\frac{\rho_{\mathrm{eq}}(x) (\partial_x\theta)^2}{2m} \right], 
\end{eqnarray}
and use the form $\theta(x,t)=\epsilon(t) x^2/2$ as a variational ansatz in the action. 

\section{Properties of the ground state wavefunction of the Generalized GPE}\label{app:ggpe-gs}
The ground state wavefunction of the generalized GPE satisfies
\begin{equation}\label{eq:ggpe} 
 -\frac{\hbar^2}{2m} \frac{d^2\psi_0}{dx^2} +[e'(\psi_0^2)+V(x)-\mu]\psi_0=0.    
\end{equation}
Multiplying Eq.(\ref{eq:ggpe}) by $\psi_0$, we find 
\begin{equation}\label{eq:ggpe-psi0}
    -\frac{\hbar^2}{2m} \frac{d}{dx}\left(\psi_0 \frac{d\psi_0}{dx}\right) + \frac{\hbar^2}{2m} \left(\frac{d\psi_0}{dx}\right)^2 +[e'(\psi_0^2)+V(x)-\mu]\psi_0^2=0,     
\end{equation}\label{eq:ggpe-dif}
while multiplying Eq.(\ref{eq:ggpe}) by $d\psi_0/dx$ and integrating, yields
\begin{equation}\label{eq:ggpe-dif}
e(\psi_0)^2 +[V(x)-\mu]\psi_0^2 - \frac{\hbar^2}{2m} \left(\frac{d\psi_0}{dx}\right)^2=\int_{-\infty}^x m\omega_0^2 y \psi_0(y)^2 dy.  
\end{equation}
Using Eq.~(\ref{eq:ggpe-dif}), we can eliminate $(V(x)-\mu)\psi_0^2$ in Eq.~(\ref{eq:ggpe-psi0}). By integrating the resulting equation from $-\infty$ to $+\infty$, we obtain 
\begin{eqnarray}
 &&\int_{-\infty}^{+\infty} dx \left[\frac{\hbar^2}{m} \left(\frac{d\psi_0}{dx}\right)^2 +e'(\psi_0^2) \psi_0^2 -e(\psi_0^2) \right] \nonumber \\
 && + \int_{-\infty}^{+\infty} dx m\omega_0^2 \int_{-\infty}^x dy y \psi_0(y)^2 =0.  
\end{eqnarray}
Using an integration by parts in the second line, we finally obtain 
\begin{equation}
    \int dx \left[\frac{\hbar^2}{m} \left(\frac{d\psi_0}{dx}\right)^2 +e'(\psi_0^2) \psi_0^2 -e(\psi_0^2) -m \omega_0^2 x^2 \psi_0(x)^2 \right]=0. 
\end{equation}
After substituting $\psi_0(x)=\sqrt{n_0(x)}$, we see that the condition above reduces to Eq.(\ref{eq:eom_varia}) with $R(t)=1, \forall t$.


\begin{thebibliography}{39}
\bibitem[{tes()}]{test}
 \bibinfo{note}{by solving the time-dependent generalized
  Gross-Pitaevskii equation, we have verified that, after initially exciting
  the mode by external radial compression of the trap, in the limit
  $|a_{1D}|\rightarrow 0$ $(\Omega_b/\omega_{ho})^2=4$.}
\end{thebibliography}
\end{document}